\documentclass[letterpaper, 10 pt, conference]{ieeeconf}  
\usepackage{graphicx}
\usepackage[latin1]{inputenc}

\IEEEoverridecommandlockouts                              
\title{\LARGE \bf
Evaluation of Alzheimer's Disease by Analysis of MR Images using
Multilayer Perceptrons and Kohonen SOM Classifiers as an Alternative
to the ADC Maps }

\author{Wellington~P.~dos~Santos,
        Ricardo~E.~de~Souza,
        and~Plínio~B.~dos~Santos~Filho
\thanks{This work was supported in part by CNPq, Brazil.}
\thanks{Wellington P. dos Santos is with the Department of Electronics and Systems of Universidade Federal de Pernambuco, 50.740-530, and the Department of Computing Systems of Universidade de Pernambuco, 50.720-001, Recife, Pernambuco, Brazil (e-mail: wellington@df.ufpe.br, wellington@dsc.upe.br)}
\thanks{Ricardo E. de Souza is with the Department of Physics of Universidade Federal de Pernambuco, 50670-901, Recife, Pernambuco, Brazil (e-mail: res@df.ufpe.br)}
\thanks{Plínio B. dos Santos Filho is with the Department of Physics of the North Carolina State University, Raleigh, North Carolina, USA}}

\begin{document}

\maketitle
\thispagestyle{empty}
\pagestyle{empty}

\begin{abstract}

Alzheimer's disease is the most common cause of dementia, yet hard to diagnose precisely without in\-va\-si\-ve techniques, particularly at the onset of the disease. This work approaches image analysis and classification of synthetic multispectral images composed by diffusion-weighted magnetic resonance (MR) cerebral images for the evaluation of cerebrospinal fluid area and measuring the advance of Alzheimer's disease. A clinical 1.5 T MR imaging system was used to acquire all images presented. The classification methods are based on multilayer perceptrons and Kohonen Self-Organized Map classifiers. We assume the classes of interest can be separated by hyperquadrics. Therefore, a 2-degree polynomial network is used to classify the original image, generating the ground truth image. The classification results are used to improve the usual analysis of the apparent diffusion coefficient map.

\end{abstract}

\section{Introduction}

Alzheimer's disease is the most common cause of dementia, both in senile and presenile individuals, observing the gradual progress of the disease as the individual becomes older \cite{ewers2006}. The major manifestation of Alzheimer's disease is the falling of the cognitive functions with gradual loss of memory, including psychological, neurological and behavioral symptoms indicating the declining of the diary life activities as a whole. Alzheimer's disease is characterized by the reduction of gray matter and the growth of cerebral sulci. However, the white matter is also affected, although the relation between Alzheimer's disease and white matter is still unknown \cite{friman2006}.

Acquisition of diffusion-wei\-ght\-ed magnetic resonance (DW-MR) images turns possible the visualization of the dilation of the lateral ventriculi temporal corni, enhancing the augment of sulci, related to the advance of Alzheimer's disease \cite{haacke1999}. Therefore, volumetrical measuring of cerebral structures is very important for the diagnosis and evaluation of the progress of diseases like Alzheimer's \cite{ewers2006}, especially the measuring of the volumes occupied by sulci and lateral ventriculi, turning possible the addition of quantitative information to the qualitative information expressed by the DW-MR images \cite{hayasaka2006}.

Usually, the evaluation of the progress of Alzheimer's disease using image analysis of DW-MR images is performed after the acquisition of at least three images of each slice of interest, acquired using the sequence spin-echo Stejskal-Tanner with different diffusion exponents, where one of the exponents is 0 s/mm$^2$, that is, a $T_2$-weighted spin-echo image \cite{haacke1999}. Then, a fourth image is calculated: the Apparent Diffusion Coefficient Map, or ADC map, where each pixel is associated to the corresponding apparent diffusion coefficient of the associated voxel: the brighter the pixels, the greater the corresponding apparent diffusion coefficients \cite{haacke1999}.

This work proposes a new approach to evaluate the progress of Alzheimer's disease: once the ADC map usually presents pixels with considerable intensities in regions not occupied by the head of patient, a degree of uncertainty can also be considered in the pixels inside the sample. Furthermore, the ADC map is very sensitive to noisy images \cite{haacke1999}. Therefore, in this case study, images are used to compose a multispectral image, where each
diffusion-weighted image is considered as a spectral band in a synthetic multispectral image.

\section{Materials and Methods}

\subsection{Diffusion-Weighted Images and ADC Maps}

The DW-MR images were acquired from the clinical images database of the Laboratory of MR Images, at the Department of Physics of Universidade Federal de Pernambuco, Recife, Brazil. The database is composed by real clinical images acquired from a clinical 1.5 T MR imaging system.

In order to compare the methods, we used 60 cerebral DW-MR images corresponding to a 70-year-old male patient with Alzheimer's disease. To perform the training of the proposed analysis, we chose the MR images corresponding to the 13th slice, showing the temporal corni of the lateral ventriculi, to furnish a better evaluation for the specialist and facilitate the correlation between the data generated by the computational tool and the \emph{a priori} knowledge of the specialist.
\begin{figure}
    \centering
        \includegraphics[width=0.25\textwidth]{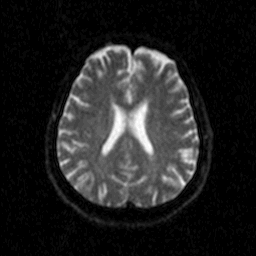}
    \caption{Axial diffusion-weighted image with exponent diffusion of 0 s/mm$^2$}
    \label{fig:epb0_3}
\end{figure}
\begin{figure}
    \centering
        \includegraphics[width=0.25\textwidth]{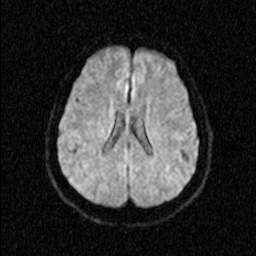}
    \caption{Axial diffusion-weighted image with exponent diffusion of 500 s/mm$^2$}
    \label{fig:epb500t_3}
\end{figure}
\begin{figure}
    \centering
        \includegraphics[width=0.25\textwidth]{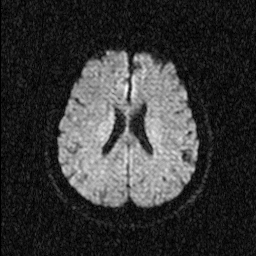}
    \caption{Axial diffusion-weighted image with exponent diffusion of 1000 s/mm$^2$}
    \label{fig:epb1000t_3}
\end{figure}

Let $f_i:S\rightarrow W$ be the set of the diffusion-weighted MR images, where $1\leq i\leq 3$, $S\subseteq \textbf{Z}^2$ is the grid of the image $f_i$, where $W\subseteq \textbf{R}$ is its codomain. The synthetic multispectral image $f:S\rightarrow W^3$ composed by the MR images of the figures \ref{fig:epb0_3}, \ref{fig:epb500t_3}
and \ref{fig:epb1000t_3} is given by:
\begin{equation}
  f(\textbf{u})=(f_1(\textbf{u}),f_2(\textbf{u}),f_3(\textbf{u}))^T,
\end{equation}
where $\textbf{u}\in S$ is the position of the pixel in the image $f$, and $f_1$, $f_2$ and $f_3$ are the diffusion-weighted MR images. Considering that each pixel $f_i(\textbf{u})$ is approximately proportional to the signal of the corresponding voxel as follows \cite{maraga2006}:
\begin{equation}
f_i(\textbf{u})=K\rho(\textbf{u})e^{-T_E/T_2(\textbf{u})}e^{-b_i
D_i(\textbf{u})},
\end{equation}
where $D_i(\textbf{u})$ is the nuclear spin diffusion coefficient measured after the $i$-th experiment, associated to the voxel mapped in the pixel in the position $\textbf{u}$; $\rho(\textbf{u})$ is the nuclear spin density in the voxel; $K$ is a constant of proportionality; $T_2(\textbf{u})$ is the transversal relaxation time in the voxel; $T_E$ is the echo time and $b_i$ is the diffusion exponent, given by \cite{haacke1999}:
\begin{equation}
b_i=\gamma^2 G_i^2 T_E^3/3,
\end{equation}
where $\gamma$ is the gyromagnetic ratio and $G_i$ is the gradient applied during the experiment $i$. Figures \ref{fig:epb0_3}, \ref{fig:epb500t_3} and \ref{fig:epb1000t_3} show images with diffusion exponents 0 s/mm$^2$, 500 s/mm$^2$ and 1000 s/mm$^2$, respectively.

The analysis of DW-MR images is often performed using the resulting ADC map $f_{\texttt{ADC}}:S\rightarrow W$, which is calculated as follows \cite{basser2002}:
\begin{equation} \label{eq:fadc}
    f_{\texttt{ADC}}(\textbf{u})=\frac{C}{b_2}\ln\left( \frac{f_1(\textbf{u})}{f_2(\textbf{u})} \right) + \frac{C}{b_3}\ln\left( \frac{f_1(\textbf{u})}{f_3(\textbf{u})} \right),
\end{equation}
where $C$ is a constant of proportionality.

Considering $n$ experiments, we can generalize equation \ref{eq:fadc} as follows:
\begin{equation} \label{eq:fadcg}
    f_{\texttt{ADC}}(\textbf{u})=\sum_{i=2}^n \frac{C}{b_i} \ln\left( \frac{f_1(\textbf{u})}{f_i(\textbf{u})} \right).
\end{equation}

Thus, the ADC map is given by:
\begin{equation}
  f_{\texttt{ADC}}(\textbf{u})=C\bar{D}(\textbf{u}),
\end{equation}
where $\bar{D}(\textbf{u})$ is an ensemble average of the diffusion coefficient $D(\textbf{u})$ \cite{fillard2006}.

\begin{figure}
    \centering
        \includegraphics[width=0.25\textwidth]{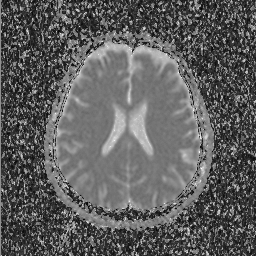}
    \caption{ADC map calculated from the three diffusion images}
    \label{fig:epb01000_3}
\end{figure}

Therefore, the pixels of the ADC map are proportional to the diffusion coefficients in the corresponding voxels. In figure \ref{fig:epb01000_3} we can see several artifacts associated to the presence of noise. In regions of the image where signal-to-noise ratio is poor (let us say, $s/n \approx 1$), the ADC map produces artifacts as consequence of the calculation of the logarithm (see equations \ref{eq:fadc} and \ref{eq:fadcg}). Therefore, the pixels of the ADC map not necessarily correspond to the diffusion coefficients but \emph{apparent} diffusion coefficients, because several pixels indicate high diffusion rates in voxels where the sample are not present or in very solid areas like, e.g., bone in the cranial box, as can be seen in figure \ref{fig:epb01000_3}. This fact can leave us to have some degree of uncertainty about the values inside the brain area.

In this work we propose an alternative to the analysis of the ADC map: the multispectral analysis of the image $f:S\rightarrow W^3$ using methods based on neural networks as an alternative that could be easily extended to other diffusion-weighted images than cerebral ones.

\subsection{Multispectral analysis using neural networks} \label{subsec_multspec}

Let the universe of classes of interest be defined as $\Omega=\{C_1,C_2,C_3\}$, $C_1$ represents cerebrospinal fluid;
$C_2$, white and gray matter, once they cannot be distinguished using diffusion images, because their diffusion coefficients are very close; $C_3$ corresponds to the image background.

For the multispectral analysis using neural networks, the inputs are associated to the vector $\textbf{x}=(x_1,x_2,x_3)^T$, where $x_i=f_i(\textbf{u})$, for $1\leq i\leq 3$. The network outputs represent the classes of interest and are associated to the vector $\textbf{y}=(y_1,y_2,y_3)^T$, where each output corresponds to the
class with the same index. The chosen decision criterion is Bayes' criterion: the output with greater value indicates the more probable class \cite{duda2001}. The training set is built using specialist knowledge at the selection of the regions of interest \cite{haykin1999}. It is composed by all the pixels of the image $f_{\texttt{ADC}}$

The synthetic multispectral image is classified using the following methods:
\begin{enumerate}
  \item \emph{Multilayer perceptron} (MLP): Initial learning rate $\eta_0=0.2$, training error $\epsilon=0.05$, maximum of 1000 training iterations, 3 inputs, 3 outputs, 2 layers, 60 neurons in layer 1 \cite{haykin1999};
  \item \emph{Kohonen SOM classifier} (KO): 3 inputs, 3 outputs, maximum of 200 iterations, initial learning rate $\eta_0=0.1$ \cite{haykin1999};
\end{enumerate}

Both methods were chosen to evaluate the behavior and the performance of a classical neural network and a clustering-based network executing the task of classification of the synthetic multispectral image. Their initial learning rates and number of iterations were empirically determined. To implement these methods, we developed a software tool called \emph{AnImed}, built using the programming environment \emph{Delphi 5}.

\subsection{Analysis of the ADC Map using Kohonen SOM Classifiers}

To make comparisons between the proposed multispectral approach and the ADC map, we performed a monospectral non-supervised classification of the ADC map using a clustering-based method \cite{li2006,bartesaghi2006}. We chose a Kohonen SOM classifier (KO-ADC) with 3 inputs, 3 outputs, maximum of 200 iterations, initial learning rate $\eta_0=0.1$.

\subsection{Noise Robustness Study}

The experiments were extended by adding artificial Gaussian noise $\xi$ to the diffusion-weighted images and the ADC map, with $\bar{\xi}=0$ and $1\%\leq \xi_{\max}\leq 20\%$, generating more 80 images. Such images were used to compose 20 synthetic multispectral images and 20 monospectral images. They were classified using the proposed methods.

\section{Results}

\begin{figure}
    \centering
        \includegraphics[width=0.25\textwidth]{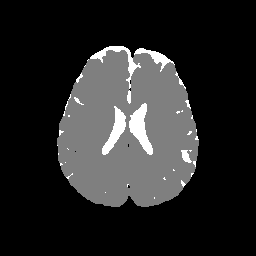}
    \caption{Ground truth image generated by the polynomial net}
    \label{fig:classPO}
\end{figure}

To evaluate objectively the classification results, we used three methods: the index $\kappa$, the \emph{overall accuracy} and the \emph{confusion matrix}. The subjective evaluation was performed by the specialist knowledge of a pathologist. Image background ($C_3$), gray and white matter ($C_2$) and cerebrospinal fluid ($C_1$) were associated to white, gray and black, respectively. Figure \ref{fig:classPO} shows the ground truth image. The regions of interest of the training set were selected using the ADC map. At the classification task, we assume the classes of interest are separable by hyperquadrics. Therefore, we chose a polynomial network with degree 2 to classify the original image and generate a ground truth image.

\begin{figure}
    \centering
        \includegraphics[width=0.25\textwidth]{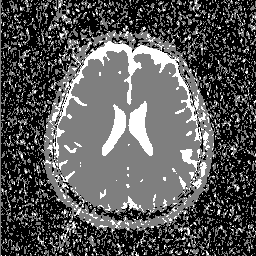}
    \caption{Classification result by the Kohonen SOM classifier using the ADC map}
    \label{fig:classKO_ADC}
\end{figure}
\begin{figure}
    \centering
        \includegraphics[width=0.25\textwidth]{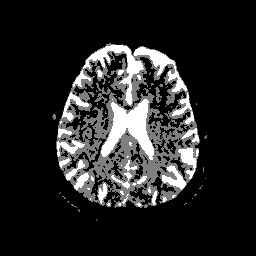}
    \caption{Classification result by the multilayer perceptron}
    \label{fig:classRNP}
\end{figure}
\begin{figure}
    \centering
        \includegraphics[width=0.25\textwidth]{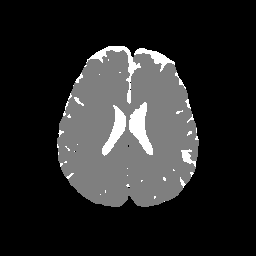}
    \caption{Classification result by the Kohonen SOM classifier}
    \label{fig:classKO}
\end{figure}
Figure \ref{fig:classKO_ADC} shows the result of the ADC map using the method KO-ADC. Figures \ref{fig:classRNP} and \ref{fig:classKO} show the results of the classification of the synthetic multispectral image composed by images \ref{fig:epb0_3}, \ref{fig:epb500t_3} and \ref{fig:epb1000t_3} using the methods MLP and KO, respectively. Table \ref{tab:ResultadosClassificacao} presents the index $\kappa$ and the overall accuracy $\phi$ for all the MR images of the entire volume (20 slices), and table \ref{tab:AreasClassificacao} shows the percentage volumes $V_1$, $V_2$ and $V_3$ occupied by the classes of interest $C_1$, $C_2$ and $C_3$, respectively, as well as the ratio between the volumes of cerebrospinal fluid and gray and white matter, simply called fluid-matter rate, expressed by $V_1/V_2$.

\begin{table}
    \caption{Percentage overall accuracy $\phi$ (\%) and index $\kappa$ by the classification methods}
  \begin{center}
  \begin{tabular} {ccccc}
  \hline
  {} & {MLP} & {KO} & {KO-ADC}\\
  \hline
  {$\phi$ (\%)} & {88.5420} & {99.3647} & {63.9809}\\
  {$\kappa$} & {0.6081} & {0.9688} & {0.2855}\\
  \hline
  \end{tabular}
  \end{center}
    \label{tab:ResultadosClassificacao}
\end{table}

\begin{table}
    \caption{Percentage volumes and fluid-matter rate by the classification methods and the ground truth image}
  \begin{center}
  \begin{tabular} {ccccc}
  \hline
  {} & {MLP} & {KO} & {KO-ADC} & {PO}\\
  \hline
  {$V_1$ (\%)} & {7.607} & {2.186} & {5.650} & {1.697}\\
  {$V_2$ (\%)} & {11.546} & {16.398} & {44.399} & {17.010}\\
  {$V_3$ (\%)} & {80.847} & {81.416} & {49.951} & {81.293}\\
  {$V_1/V_2$} & {0.659} & {0.133} & {0.127} & {0.100}\\
  \hline
  \end{tabular}
  \end{center}
    \label{tab:AreasClassificacao}
\end{table}

\begin{figure}
    \centering
        \includegraphics[width=0.40\textwidth]{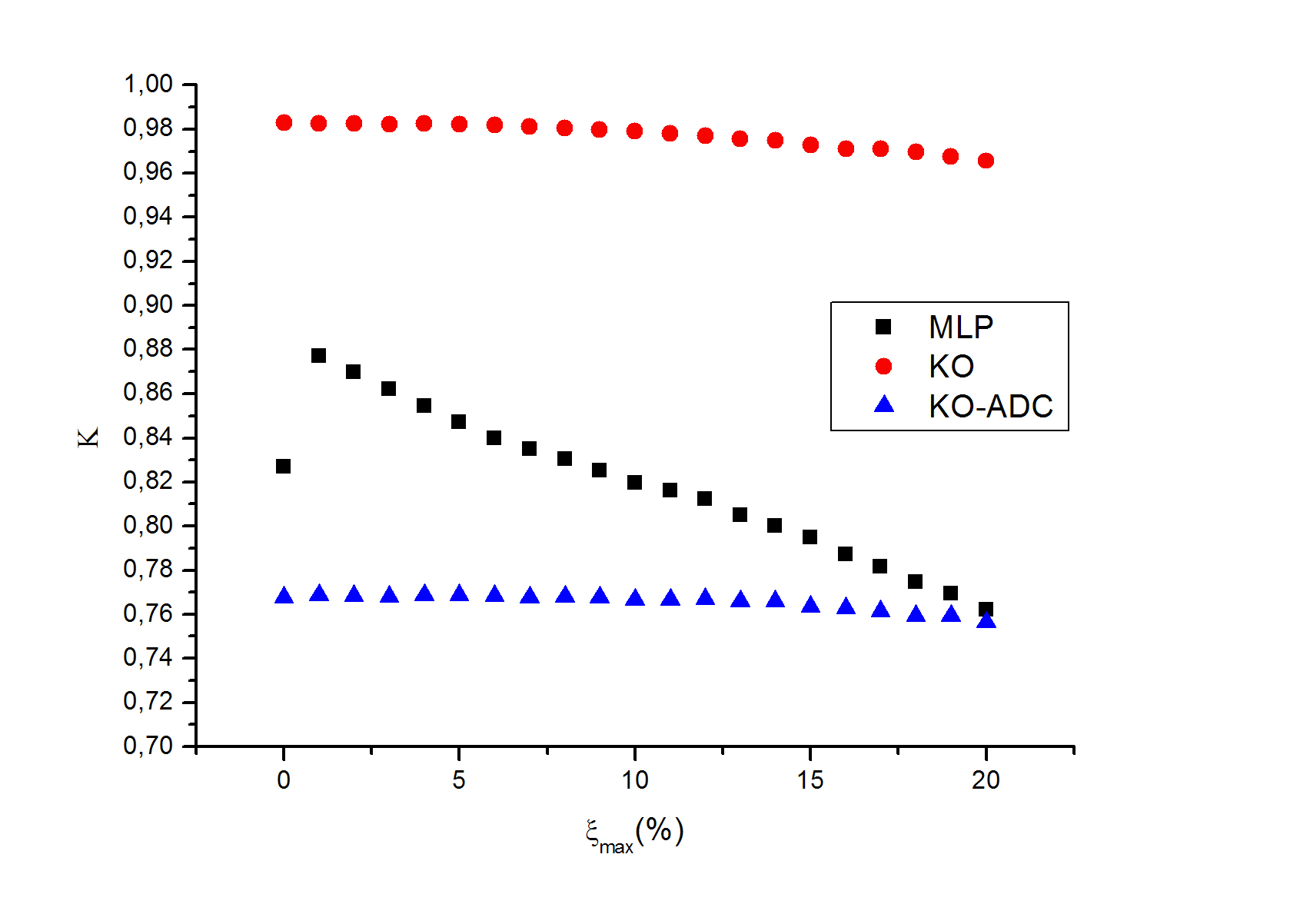}
    \caption{Behavior of the index $\kappa$ with the variation of maximum Gaussian noise $\xi_{\max}$}
    \label{fig:Graph1}
\end{figure}

Figure \ref{fig:Graph1} shows the behavior of the index $\kappa$ versus the maximum Gaussian noise $\xi_{\max}$ for the methods MLP, PO, KO, and KO-ADC, applied to the 13th slice, used to mount the training set of the polynomial net.

\section{Discussion and Conclusions}

From table \ref{tab:ResultadosClassificacao} we can see that the multispectral approach, with index $\kappa$ of 0.6081 and 0.9688 for MLP and KO classifiers, respectively, is superior to the analysis of the ADC map, with index $\kappa$ of 0.2855. Such results are confirmed when we observe the classification results of the multispectral approach on the 13th slice, in figures \ref{fig:classRNP}, \ref{fig:classPO} and \ref{fig:classKO}, and compare to the result of the analysis of the ADC map in figure \ref{fig:classKO_ADC}, where we can see several areas out of the sample and in the cranial box wrongly marked as cerebrospinal fluid and matter.

The classification results by MLP method (figure \ref{fig:classRNP}) overestimated the area occupied by cerebral fluid. When this result is compared to the diffusion image with diffusion exponent of 0 (figure \ref{fig:epb0_3}), we can see that left and right ventriculi are separated. Furthermore, the sulci were also overestimated, which could leave the specialist to evaluate this Alzheimer's case as more advanced than really it is. The fluid-matter rate ($V_1/V_2$) equals 0.659 for the multilayer perceptron, greater than the results obtained from the analysis of the ADC map, with $V_1/V_2 = 0.127$ and the rate obtained by the polynomial net, with $V_1/V_2 = 0.100$.

The behavior of the index $\kappa$, according to the maximum Gaussian noise $\eta_{\max}$, demonstrated that the KO method is more robust than the MLP classification. The index $\kappa$ of the MLP method rapidly decays as the maximum level of noise increases. Consequently, we can discharge the result by the MLP method and consider a good estimation for the real fluid-matter rate as the result obtained by KO.

The multispectral classification of diffusion-weighted MR images furnishes a good alternative to the analysis of the ADC map, consisting on a very reasonable mathematical tool useful to perform qualitative and quantitative analysis of the progress of Alzheimer's disease for the medical specialist.


\bibliographystyle{unsrt}
\bibliography{arq_bib}

\end{document}